\documentclass[%
 reprint,
 amsmath,amssymb,
 aps,
prl,
]{revtex4-2}

\usepackage{xcolor}
\usepackage{graphicx}
\usepackage{dcolumn}
\usepackage{bm}

\begin{document}

\preprint{APS/123-QED}

\title{
Elasticity, stability, and quasi-oscillations of cell-cell junctions\\ in solid confluent epithelia
}

\author{Cl\'ement Zankoc}
\affiliation{Jo\v zef Stefan Institute, Jamova 39, SI-1000 Ljubljana, Slovenia}


\author{Matej Krajnc}
 \email{matej.krajnc@ijs.si}
\affiliation{Jo\v zef Stefan Institute, Jamova 39, SI-1000 Ljubljana, Slovenia}%



\begin{abstract}
Macroscopic properties and shapes of biological tissues depend on the remodelling of cell-cell junctions at the microscopic scale. We propose a theoretical framework that couples a vertex model of solid confluent tissues with the dynamics describing generation of local force dipoles in the junctional actomyosin. Depending on the myosin-turnover rate, junctions either preserve stable length or collapse to initiate cell rearrangements. We find that noise can amplify and sustain transient oscillations to the fixed point, giving rise to quasi-periodic junctional dynamics. We also discover that junctional stability is affected by cell arrangements and junctional rest tensions, which may explain junctional collapse during convergence and extension in embryos.

\end{abstract}

\maketitle


\section*{Introduction}

In confluent tissues, the adjacent cells adhere to one another through narrow joints known as the adherens junctions. These regions are rich with protein complexes, which govern cell-cell adhesion and couple cells' cytoskeletons~\cite{harris10,vasquez16}. Forces transmitted along the junctions play a major role in morphogenesis. In particular, contractile tensions generated in the actomyosin~\cite{lecuit07,lecuit11,murrell15} drive various types of cell deformations and tissue-scale movements~\cite{fernandez-gonzalez09,heisenberg20}. For instance, the actomyosin network collapses during cell ingression and oscillates during dorsal closure in {\it Drosophila}~\cite{simoes17,aroshan13,martin09,lin17,lo18}, directed junctional collapse drives convergence and extension in the early {\it Drosophila} embryo~\cite{bertet04,rauzi10,staddon20,stern20}, whereas junctional fluctuations establish the arrangement of cells in {\it Drosophila} pupal notum~\cite{curran17} and fluidize the tissue during vertebrate body axis elongation~\cite{mongera18}. 

Recent measurements of junctional dynamics during tissue remodelling revealed that the rate of junctional collapse often increases with contraction~\cite{curran17,stern20}. This suggests a positive feedback loop between junctional contraction and generation of junctional tension. Apart from driving collapse of the actomyosin, such a feedback loop could establish oscillatory dynamics as previously proposed within a generic theoretical framework of active contractile elements~\cite{dierkes14}.

To explore these dynamics in the context of confluent tissues, we develop a vertex model with a feedback loop between cell-scale junctional contractions and generation of force dipoles at the level of the junctional actomyosin. We find that the nonlinear elastic response of solid tissues to local force dipoles does not satisfy conditions to yield a limit cycle of junctional oscillations. Nevertheless, the variety of dynamical regimes remains rich and includes junctions that can either sustain stable length or collapse. One of our main findings shows that junctional noise establishes quasi-periodic junctional dynamics. We also discover that junctional stability depends on cell arrangements as well as on distribution of junctional rest tensions, which could be relevant for the active junctional remodelling during morphogenesis~\cite{bertet04,rauzi10}.

\section*{Materials and Methods}

\subsection*{The model}
The tissue is represented by a planar polygonal network of cell-cell junctions parametrized by the positions of cell vertices $\boldsymbol r_i=\left (x_i,y_i\right )$~\cite{fletcher14,alt17,barton17}. Forces on vertices $\boldsymbol F_i$ are assumed conservative such that $\boldsymbol F_i=-\nabla_iW$, where $W=\kappa\sum_{k}(A_k-A_0)^2+\sum_{ij}\gamma_{ij}l_{ij}$ is the total potential energy of the system and $\nabla_i=\left (\partial/\partial x_i,\partial/\partial y_i\right )$. The first sum in the energy goes over all cells and describes cell-area elasticity ($A_k$ and $A_0$ being the actual and the preferred cell areas, respectively), whereas the second sum goes over all junctions and describes the line energy due to adhesion and cortical tension~($l_{ij}$ being the junctional length); indices $i$ and $j$ denote head and tail vertices of the junctions. For simplicity, we disregard the usual cell-perimeter-squared energy term~\cite{farhadifar07,bi15}, which is not needed to describe solid tissues studied here. Due to strong friction, the motion of vertices is overdamped:
\begin{equation}
    \label{eq1}
    \eta\dot{\boldsymbol r}_i=\boldsymbol F_i=-2\kappa\sum_{k}(A_k-A_0)\nabla_iA_k-\sum_{j}\gamma_{ij}\nabla_il_{ij}\>,
\end{equation}
where $\eta$ is the friction coefficient. The tension $\gamma_{ij}(t)=\gamma_0+\Delta\gamma_{ij}(t)$, where $\gamma_0$ and $\Delta\gamma_{ij}(t)$ are the average (rest) tension and the time-depending part, respectively. To ensure stability, $\gamma_0$ is assumed positive, which implies that the cortical tension dominates over the adhesion~\cite{curran17,krajnc20}. From now on we use dimensionless quantities by choosing $\sqrt{A_0}$, $\tau_0=(\eta\gamma_0)^{-1}$, and $\gamma_0$ as the units of length, time, and tension, respectively, and rescaling the modulus as $\kappa A_0^{3/2}/\gamma_0\to\kappa$.

\begin{figure}[t!]
    \includegraphics[]{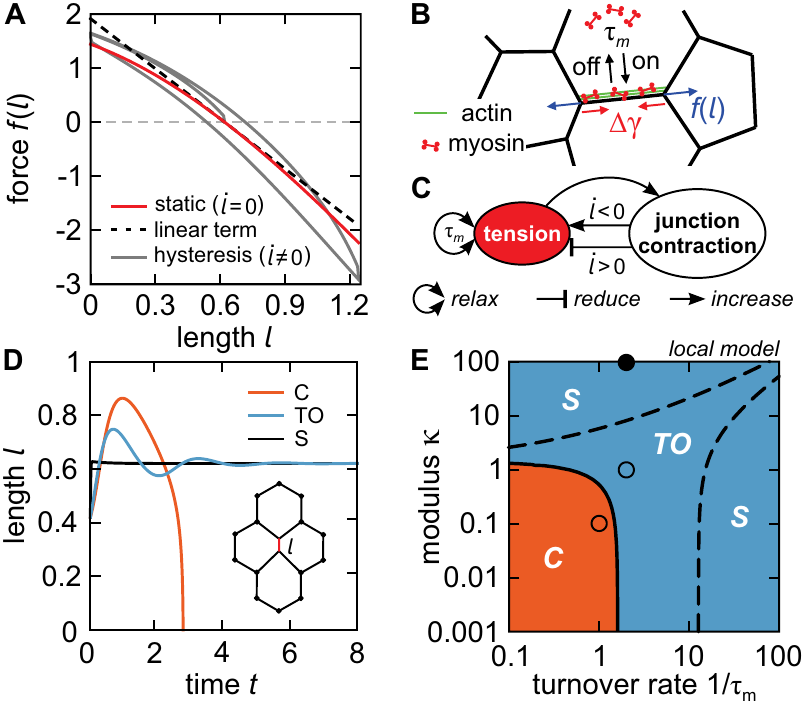}
    \caption{(A)~Static force-extension relation~(red curve) and hysteresis~(grey curve) for a single junction in the honeycomb cell tiling. (B)~Schematic of the interplay between the elastic restoring force $f(l)$ and active tension $\Delta\gamma$ generated by the junctional myosin. (C)~The feedback loop between junctional contraction and tension. (D)~Junction length vs. time in the local model. (E)~Phase diagram of local model exhibits stable~(S), collapse~(C), and stable dynamics with transient oscillations to the fixed point~(TO). Red, blue, and black curves~(panel D) and circles (panels E) correspond to $(1/\tau_m,\kappa)=(1,0.1)$, $(2,1)$, and $(2,100)$, respectively.}
    \label{F1}
\end{figure}

Tissues with constant line tensions ($\gamma_{ij}={\rm const.}$) behave as passive elastic materials capable of sustaining shear stresses~\cite{farhadifar07,staple10}. To examine their response to a locally applied force dipole, we assume an excess line tension~$\Delta\gamma$ on a single junction. This deforms the junction away from its rest length $l_0$, and the elastic deformation propagates into the bulk~(Supporting Materials and Methods, Fig.~S1A). Force balance implies $0=-2\Delta\gamma-f(l)$, where $f(l)$ is the elastic restoring force, which is well described by $f(l)=\sum_{m=1}^3\alpha_m\left ( l-l_0\right )^m$ with $\alpha_2/\alpha_1=0.33$ and $\alpha_3/\alpha_1=-0.10$ for a regular honeycomb cell tiling~(Fig.~\ref{F1}A and Supporting Materials and Methods, Fig.~S1B,C). The significant contribution of the second-order term yields softening and stiffening behaviors for compression and stretch, respectively~(Fig.~\ref{F1}A). Furthermore, due to friction, $f(l)$ displays hysteresis at non-zero rates of change of junctional length~(Fig.~\ref{F1}A).

Tensions change on time scales associated with the dynamics of the underlying actomyosin~(Fig.~\ref{F1}B). We assume a linear relation between junctional tension and myosin concentration~[defined by a number of molecular motors per junction length $c_{ij}(t)=N_{ij}/l_{ij}$]: $c_{ij}(t)=\alpha\gamma_{ij}(t)$, where $\alpha$ is a constant proportionality factor. The total rate of change of the myosin concentration $\dot{c}_{ij}=\dot{N}_{ij}/l_{ij}-N_{ij}\dot{l}_{ij}/l_{ij}^2$, where the first and the second term describe changes of the number of motors at a fixed junction length and changes of the junction length at a fixed number of motors, respectively. In particular, the motor-actin binding and undbinding contribute a time scale $\tau_m$ and are described by $\dot{N}_{ij}=(-N_{ij}+c_0l_{ij})/\tau_m$, where $c_0=\alpha\gamma_0$ is the ambient myosin concentration. In turn, $\dot{l}_{ij}$ can be explicitly written in terms of the forces acting at the vertices as $\dot l_{ij}=\boldsymbol r_{ij}\cdot\boldsymbol F_{ij}/l_{ij}$, where $\boldsymbol r_{ij}=\boldsymbol{r}_j-\boldsymbol{r}_i$ and $\boldsymbol F_{ij}=\boldsymbol{F}_j-\boldsymbol{F}_i$. Overall, this yields a deterministic dynamic equation for tension fluctuations:
\begin{equation}
    \label{eq2}
    \dot{\Delta\gamma_{ij}}=-\frac{1}{\tau_m}\Delta\gamma_{ij}-\frac{\gamma_{ij}\boldsymbol r_{ij}\cdot\boldsymbol F_{ij}}{l_{ij}^2}\>,
\end{equation}
where the first term describes tension relaxation due to myosin turnover, whereas the second term describes a coupling between tension dynamics and mechanics of the vertex model. This coupling gives rise to a feedback loop between generation of tension and junctional contractions~(Fig.~\ref{F1}C). In contrast to some previous studies, which describe similar feedback mechanisms using specific chemomechanical models~\cite{lin17,siang18,staddon20}, here the feedback follows directly from the relation between junctional tension and myosin concentration.

\section*{Results}

\subsection*{Local model}

First, we apply the model on a quartet of hexagonal cells~(inset to Fig.~\ref{F1}D). We derive the equation of state for the central junction, which reads
\begin{equation}
    f(l)=-3\kappa l_0^3\left (\frac{l}{l_0}-1\right )-2-\frac{4(l/l_0-2)}{\sqrt{(l/l_0)^2-4l/l_0+7}}\>,
\end{equation}
where $l_0=3^{-3/4}\sqrt{2}\approx 0.62$ is the side length of a regular hexagon with unit area. Interestingly, despite its simplicity, the local model exhibits similar elasticity than the full tissue--including softening/stiffening behavior upon junction compression/stretch~(Supporting Materials and Methods, Sec.~I and Fig.~S2A,B). Next, we simulate the dynamics described by~Eqs.~(\ref{eq1}) and (\ref{eq2}), starting in a configuration away from the fixed point $(l,\gamma)_0=(l_0,1)$. We find three types of behavior: The junction either (i)~converges directly back to the fixed point~(state S), (ii)~undergoes damped transient oscillations before reaching the fixed point~(state TO), or (iii)~collapses its length to 0~(state C)~(Fig.~\ref{F1}D). The exact time course of junction length, $l(t)$, depends  on the initial conditions.

A linear stability analysis~(Supporting Materials and Methods, Sec.~II) reveals the analytical condition for the Hopf bifurcation:
\begin{equation}
    \label{eq3}
    \kappa^*=\frac{1}{3l_0^3}-\frac{1}{3l_0^2\tau_m}\>.
\end{equation}
In particular, the junction is stable for $\kappa>\kappa^*$ and unstable for $\kappa<\kappa^*$~(Fig.~\ref{F1}E). Transient oscillations appear within the stable regime at $\kappa_{-}<\kappa<\kappa_{+}$, where $\kappa_{\pm}=1/(3l_0^3)+1/(3l_0^2\tau_m)\pm(1/3l_0^2)\sqrt{10/(l_0\tau_m)}$~(Fig.~\ref{F1}E). Importantly, the first Lyapunov coefficient shows that the Hopf bifurcation is always subcritical, meaning that no parameter values yield a stable limit cycle that would describe periodic oscillations like those previously observed in similar dynamical models~\cite{lin17,lo18,dierkes14}~(Supporting Materials and Methods, Sec.~III). 

\begin{figure}[htb!]
    \begin{center}
    \includegraphics[]{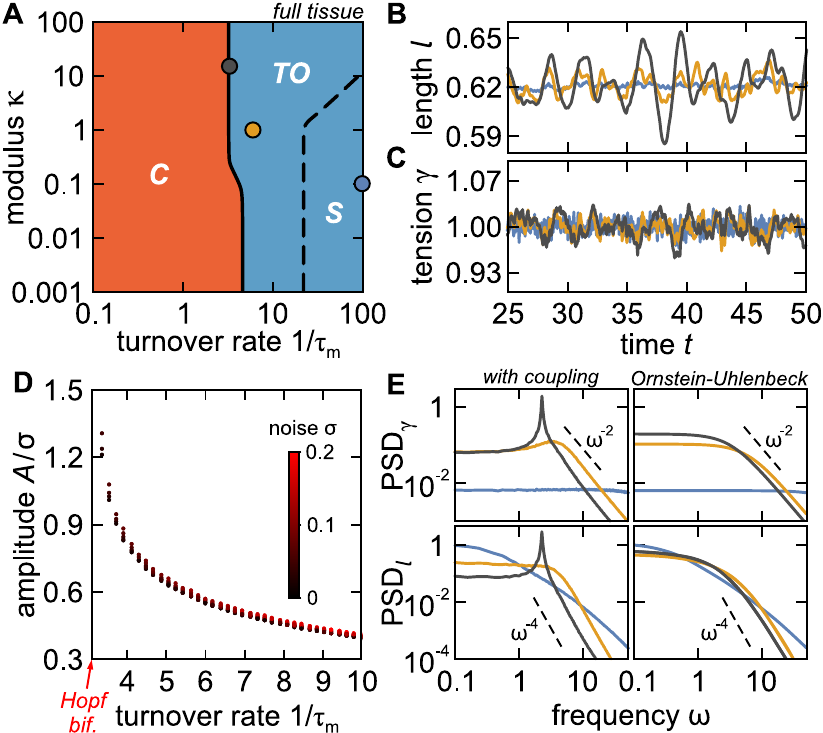}
    \caption{(A)~Phase diagram of the full tissue exhibits stable~(S), collapse~(C), and transient oscillations to the fixed point~(TO). Eigenvalues of the Jacobian were evaluated at $(1/\tau_m,\kappa)$-pairs on an equidistant grid of $100\times100$ points. (B,C)~Junctional length~(B) and tension~(C) vs. time for examples with quasi-oscillations~(grey and orange curves) and stochastic fluctuations~(blue). (D)~Amplitude of length fluctuations vs. myosin turnover rate $1/\tau_m$ for $\sigma=0.01-0.2$ (black-to-red color scheme). Raw (non-collapsed) data are shown in~Supporting Materials and Methods, Fig.~S4. (E)~Power spectral densities of length and tension fluctuations. Grey, orange, and blue circles~(panel A) and curves (panels B, C, and E) correspond to $(1/\tau_m,\kappa)=(3.25,15)$, $(6,1)$, and $(100,0.1)$, respectively, whereas $\sigma=0.01$.}
    \label{F2}
    \end{center}
\end{figure}

Next, to search for the various types of junctional behaviors in the full-tissue setting, we apply our model to the honeycomb cell tiling and numerically preform the linear stability analysis~(Supporting Materials and Methods, Sec.~IV). We identify the same dynamical regimes as in the local model~(Fig.~\ref{F2}A). However, in contrast to the local model, the collapse regime extends to large $\kappa$ values~(Fig.~\ref{F2}A). This is because collective cell deformations in the full-tissue model allow preservation of cell areas even upon junction collapse.

\subsection*{Quasi-oscillations}

Binding and undbinding of myosin are stochastic processes, which cause stochastic fluctuations of junctional tensions~\cite{curran17,krajnc18,okuda19,krajnc20}. To explore the role of noise, we add an extra term to the equation for tension, which now reads $\dot{\gamma}_{ij}=\Delta\dot{\gamma}_{ij}+\sqrt{2\sigma^2/\tau_m}\>\xi_{ij}$. Here the first term obeys Eq.~(\ref{eq2}) like before, whereas the second term describes the white noise with long-time variance $\sigma^2$ and $\langle\xi_{ij}(t)\rangle=0$, $\langle\xi_{ij}(t)\xi_{kl}(t')\rangle=\delta_{ik}\delta_{jl}\delta(t-t')$. Note that in the absence of the coupling term in~Eq.~(\ref{eq2}), our stochastic tension dynamics reduces to the classical Ornstein-Uhlenbeck~(OU) process, which was previously used to describe stochastic junctional fluctuations~\cite{curran17,krajnc20}.

We simulate the stochastic dynamics in a honeycomb cell tiling and find that while in the stable (S) regime, length (and tension) fluctuations are noisy~(blue curves in Fig.~\ref{F2}B and C), the movements become more regular in the regime of transient oscillations~(TO) and, in contrast to the deterministic case, manage to sustain a well-defined amplitude~(orange and grey curves in Fig.~\ref{F2}B and C). These movements in fact correspond to the transient oscillations, which eventually die out in the purely deterministic case~(Fig.~\ref{F1}D), but get amplified~(Fig.~\ref{F2}D) and become sustained indefinitely in the presence of the noise. This gives rise to a quasi-periodic trajectory or the so-called quasi-cycle, which arises in dynamical systems in the presence of the noise when the Jacobian matrix has complex eigenvalues with strictly negative real parts~\cite{boland08,mckane08,zankoc17}. We refer to these junctional movements as quasi-oscillations.

Next, we examine the observed dynamics in the Fourier space. The quasi-oscillations give rise to peaks in power spectral densities~(PSD) of tension- and length fluctuations, PSD$_\gamma$ and PSD$_l$, respectively. These peaks get dominated by the noise when moving away from the bifurcation point~(Fig.~\ref{F2}E and Supporting Materials and Methods, Sec.~V). In fact, in the limit $1/\tau_m\to\infty$ where the relative contribution of the coupling term in the tension dynamics~[Eq.~(\ref{eq2})] becomes negligible compared to the relaxation term, the PSDs agree with the model in which tensions obey a pure OU process~(blue curves in Fig.~\ref{F2}E).

In confluent tissues, junctions are interconnected and so need to synchronize their quasi-oscillations. Since three junctions meet at each vertex, junction networks are geometrically frustrated, which can lead to nontrivial spatial patterns of cell deformations~(Supporting Materials and Methods, Sec.~VI).

\subsection*{Inhomogeneous tissues}

\begin{figure}[htb!]
     \begin{center}
    \includegraphics[]{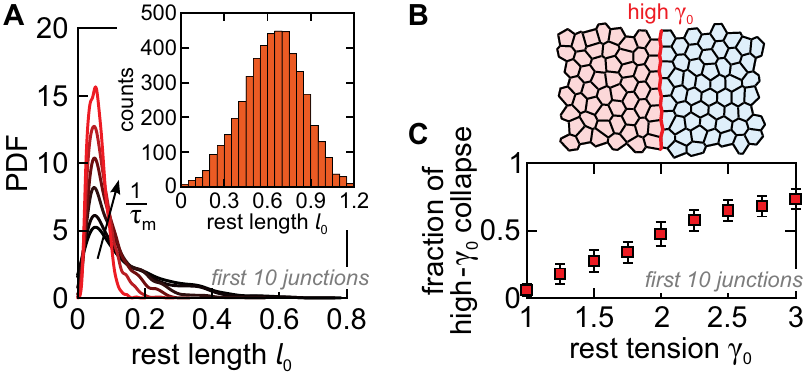}
    \caption{(A)~Probability distribution function~(PDF) of the rest length of first 10 collapsing junctions for $1/\tau_m=0.01,\>1,\>3,\>5,\>7.4$ and $10$~(black-to-red color scheme). Inset: Distribution of rest lengths in $100$ analyzed tissues. (B)~Tissue with a supracellular myosin cable of junctions with $\gamma_0\geq 1$ (red junctions). The rest tensions on black junctions equal 1. (C)~Fraction of high-$\gamma_0$ collapsed junctions versus $\gamma_0$.}
    \label{F4}
    \end{center}
\end{figure}

The stability condition of our local model [Eq.~(\ref{eq3})] suggests that disorder of cell packing may affect junctional stability. Indeed, it can be recast as $l_0>l_0^*$, where the $l_0^*$, is the critical rest length for collapse. Since the rest lengths are distributed in disordered tissues~(inset to Fig.~\ref{F4}A), there might exist a fraction of junctions that are shorter than $l_0^*$. These junctions would necessarily collapse and trigger cell rearrangements. To test this possibility, we examine an ensemble of 100 disordered tissues~(Supporting Materials and Methods, Sec.~VII). Starting close to the fixed point (i.e., all junctions at their rest lengths and all tensions equal 1), we simulate the dynamics~[Eqs.~(\ref{eq1}) and~(\ref{eq2})] at fixed $\kappa=15$ and record the rest lengths of the first 10 collapsing junctions. Here, each junctional collapse initiates a T1 transition, which is performed as soon as the junction length drops below $0.01$. The length of the newly created junction is set to 0.001, whereas its tension is reset to $\gamma_0$. Unlike in the honeycomb lattice where junctions at $\kappa=15$ collapse only for $1/\tau_m<3.2$~(Fig.~\ref{F2}A), in disordered tissues we find short junctions collapsing even at higher $1/\tau_m$ values~(Fig.~\ref{F4}A). This confirms that the distribution of rest lengths in disordered tissues importantly affects local junctional stability. We note in passing that frequent collapses of short junctions drive partial ordering of disordered tissues~(Supporting Materials and Methods, Fig.~S6 and Videos.~S1 and S2).

Finally, the dimensional form of the stability condition, $\kappa^*>\gamma_0/(3 l_0^3)-\eta/(3 l_0^2 \tau_m)$, suggests that the rest junctional tension $\gamma_0$ may also affect junctional stability. In particular, junctions with higher $\gamma_0$ are expected to have an increased critical rest length $l_0^*$, possibly resulting in their collapse. We test this prediction by analyzing a tissue that contains a supracellular cable of enriched junctional myosin, which increases the rest tensions on the corresponding junctions compared to other junctions~(Fig.~\ref{F4}B). Without the contraction-tension coupling~[Eq.~(\ref{eq2})], this cable would be stable despite being under higher tension, since every vertex within the cable is acted upon by a pair of equal but opposite forces. However, as predicted by the stability condition, the coupling indeed affects the stability of the junctions that are under higher rest tension, making them more succesptible for collapse. To show this, we record the first 10 collapsing junctions and find that the fraction of those with high $\gamma_0$ increases with $\gamma_0$~(Fig.~\ref{F4}C).

\section*{Discussion}

We studied a mechanical model of tissues with a feedback loop between junctional contractions and the dynamics of junctional tensions~(Fig.~\ref{F1}). In particular, we used a previously proposed description of force generation at the level of the actomyosin~\cite{dierkes14} and combined it with the vertex model of solid confluent tissues, which provided a faithful representation of the elasticity underlying the response of solid tissues to local force dipoles at the junctions. While nonlinearities in this system do not meet the conditions to yield a stable limit cycle of junctional oscillations~(Fig.~\ref{F1}A), we discovered that junctional noise can amplify and sustain quasi-periodic junctional dynamics at biologically relevant myosin-turnover rates~[Fig.~\ref{F2} and Ref.~\cite{curran17}]. Importantly, this dynamical regime does not even require nonlinear elasticity and may thus be more common than the limit cycle of periodic junctional oscillations. Another important result of our work highlights the role of cell arrangements and the distribution of rest tensions within the tissue for the junctional stability~(Fig.~\ref{F4}). Both effects may be present during convergence and extension in {\it Drosophila} embryo, where the so-called parasegmental boundaries, enriched with the junctional myosin, frequently collapse their junctions~\cite{bertet04,rauzi10,rauzi20}.

An interesting future direction would be to employ the Area- and Perimeter-Elasticity vertex model~\cite{farhadifar07,bi15} and use it to explore tissues that are closer to the solid-fluid transition. These tissues are associated with highly nonlinear elasticity~\cite{staple10,bi15,sahu20}, which could, in contrast to our model, yield a stable limit cycle of junctional oscillations~\cite{dierkes14}. In turn, this could lead to spontaneous organization of junctions into groups of locally synchronized oscillators. Furthermore, the role of correlations between junctions needs to be further investigated. In particular, these correlations can appear because individual junctions are under tension exerted by the actomyosins from {\it two} adjacent cells and because cell membranes belonging to the same cell share a common pool of molecular motors. These effects would provide additional sources of coupling between junctions, which could significantly affect their correlated movements. Finally, an alternative model could also assume that junctional tension is proportional to the number of myosin motors rather than to their concentration.
\section*{Acknowledgments}
We thank Primo\v z Ziherl, Jan Rozman, Tomer Stern, Guillaume Salbreux, and Fabio Staniscia for critical reading of the manuscript. We acknowledge the financial support from the Slovenian Research Agency (research project No. Z1-1851 and research core funding No. P1-0055).


\end{document}


\preprint{APS/123-QED}

\title{
Elasticity, stability and quasi-oscillations of cell-cell junctions\\ in solid confluent epithelia: Supplemental Material
}

\author{Cl\'ement Zankoc}


\author{Matej Krajnc}



\begin{abstract}

\end{abstract}

\maketitle
%
%
%

%
\begin{figure}[htb!]
    \begin{center}
    \includegraphics[width=85mm]{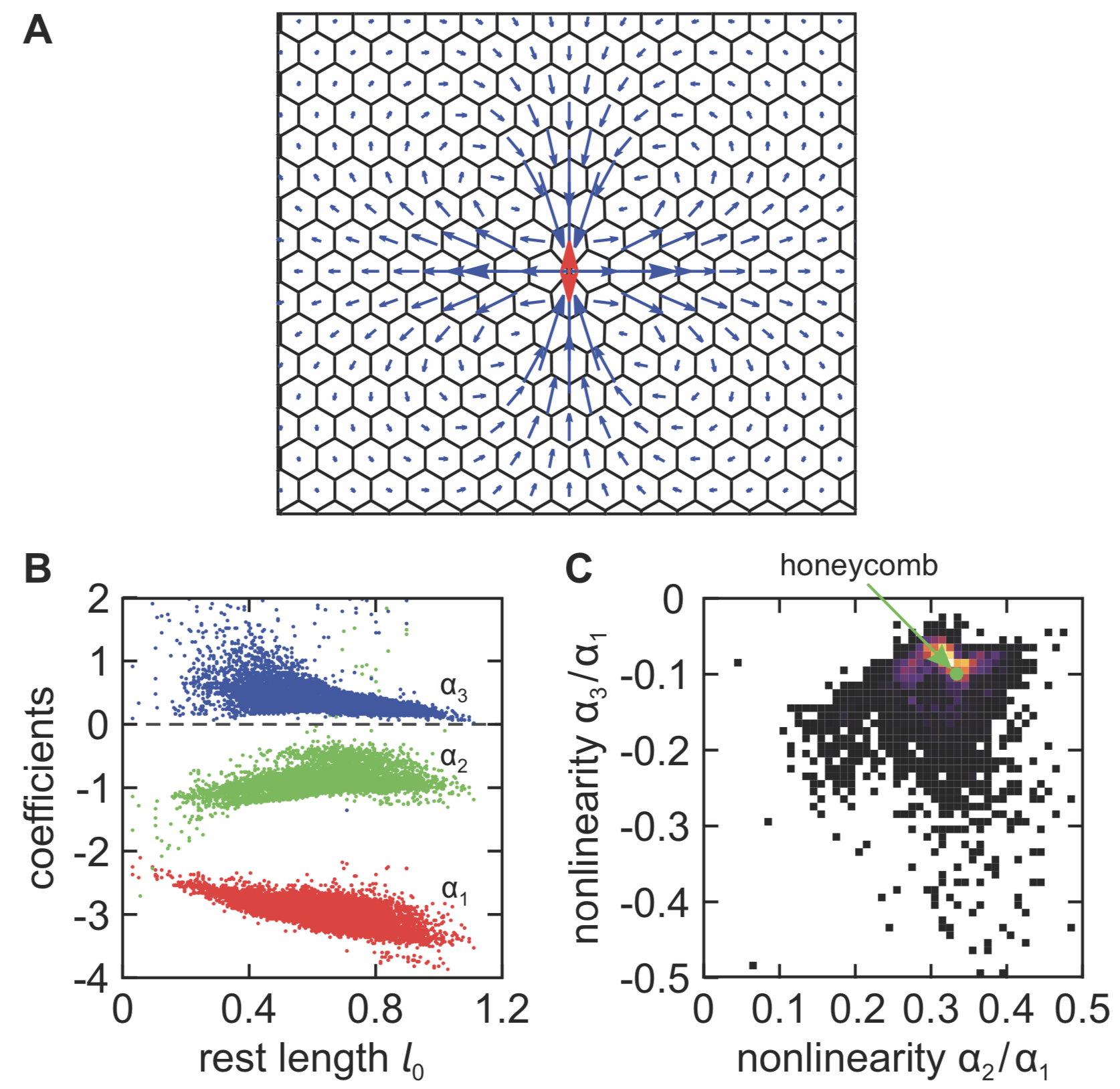}
    \caption{(A)~Deformation field due to a local force dipole~(red arrows) in a honeycomb cell tiling. (B)~Coefficients $\alpha_i$ vs. rest length $l_0$ from force-extension relations $f(l)=\sum_{m=1}^3\alpha_m(l-l_0)^m$ for junctions extracted from disordered tissues. (C)~The distribution of second- and third-order non-linear coefficients in disordered tissues. The green circle in panel B denotes the pair of nonlinear coefficients corresponding to the state equation of an edge in the honeycomb lattice of cells.}
    \label{FS1}
    \end{center}
\end{figure}
%
\section{Equation of state}
\begin{figure}[t!]
    \begin{center}
    \includegraphics[width=85mm]{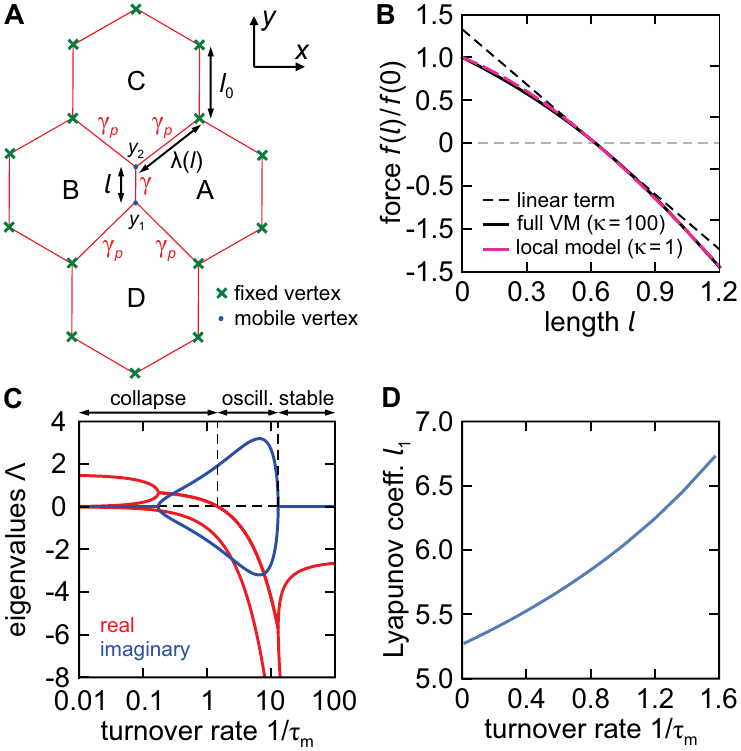}
    \caption{(A)~Schematic of the local model. Vertices denoted by green crosses are fixed, whereas the blue vertices are allowed to move along the $y$ direction. Variables of the model are the length of the central edge $l$, and tensions $\gamma$ and $\gamma_p$ on the central and 4 peripheral edges, respectively. The lengths of peripheral edges $\lambda(l)$ can be explicitly calculated from $l$. (B)~Force-extension curves of the full vertex model~(black curve) and the local model~(purple curve). Dashed line shows the linear part of the elastic restoring force. (C)~Eigenvalues of the Jacobian versus turnover rate $1/\tau_m$ at fixed $\kappa=0.1$. (D)~First Lyapunov coefficient as a function of $1/\tau_m$.}
    \label{FS2}
    \end{center}
\end{figure}

We treat a quartet of hexagonal cells in which the central pair of vertices with positions $\boldsymbol r_1=(0,y_1)$ and $\boldsymbol r_2=(0,y_2)$ move symmetrically with respect to the $x$-axis (i.e., $y_1=-y_2$), whereas all the other vertices are fixed~(Fig.~\ref{FS2}A). While such vertex displacements preserve the total area of this four-cell neighborhood, the areas of individual cells $A_A$, $A_B$, $A_C$, and $A_D$ do change with vertex displacements. Nevertheless, the system possesses two axes of mirror symmetry, meaning that $A_A = A_B$, $A_C = A_D$, whereas the length of all four edges adjacent to the central edge,
\begin{equation}
    \lambda(l) =  \sqrt{\left( \frac{\sqrt{3}}{2} l_0 \right)^2 + \left( \frac{l}{2} - l_0 \right)^2}\>. 
\end{equation}
We refer to these edges as peripheral edges and to preserve the symmetry, we assume that they are under the same tension $\gamma_p=\gamma_p(t)$; tension on the central edge is diffrerent from that at peripheral edges and is denoted by $\gamma$.

The total energy of the system is given by
\begin{equation}
    W = \kappa\sum_{\alpha} \left( A_{\alpha} - A_0 \right)^2 + \sum_{ij}\gamma_{ij} l_{ij}\>.
\end{equation}
We are interested in the dynamics of the central edge length $l=y_2-y_1$. Therefore, we need to first derive the dynamics equations for $y_1$ and $y_2$, which obey $\eta{\rm d}y_1/{\rm d}t=-\partial W/\partial y_1$ and $\eta{\rm d}y_2/{\rm d}t=-\partial W/\partial y_2$, respectively. In turn, this gives
\begin{equation}
    \eta \frac{\mathrm{d}y_1}{\mathrm{d}t} = -2\kappa\sum_\alpha\left(A_\alpha-A_0\right) \frac{\partial A_\alpha}{\partial y_1} - \gamma \frac{\partial l}{\partial y_1} -  2 \gamma_p \frac{\partial \lambda(l) }{\partial y_1} 
\end{equation}
and
\begin{equation}
    \eta \frac{\mathrm{d}y_2}{\mathrm{d}t} = -2\kappa\sum_\alpha\left(A_\alpha-A_0\right) \frac{\partial A_\alpha}{\partial y_2} - \gamma \frac{\partial l}{\partial y_2} -  2 \gamma_p \frac{\partial \lambda(l) }{\partial y_2} \>.
\end{equation}
Next, we express cell areas with respect to $l$ as follows:
\begin{equation}
    A_A = A_B= \frac{\sqrt{3}}{4}l_0\left (5l_0+l\right )
\end{equation}
and
\begin{equation}
    A_C=A_D = \frac{\sqrt{3}}{4}l_0\left (7l_0 - l\right )\>.
\end{equation}
The derivatives of the areas with respect to $y_1$ and $y_2$ read
\begin{equation}
    \frac{\partial A_A}{\partial y_1} = \frac{\partial A_B}{\partial y_1} = - \frac{\sqrt{3}}{4} l_0\>,
\end{equation}
\begin{equation}
    \frac{\partial A_A}{\partial y_2} = \frac{\partial A_B}{\partial y_2} = \frac{\sqrt{3}}{4} l_0\>,
\end{equation}
\begin{equation}
    \frac{\partial A_C}{\partial y_1}=\frac{\partial A_D}{\partial y_1} = \frac{\sqrt{3}}{4} l_0\>,
\end{equation}
and
\begin{equation}
    \frac{\partial A_C}{\partial y_2}= \frac{\partial A_D}{\partial y_2} = - \frac{\sqrt{3}}{4} l_0 \>,
\end{equation}
whereas the derivatives of the lengths are $\partial l/\partial y_1 = - 1$, $\partial l/\partial y_2 = 1$, and
\begin{equation}
    \frac{\partial \lambda(l)}{\partial y_1} =-\frac{\partial \lambda(l)}{\partial y_2}= -\frac{l - 2l_0}{2\lambda(l)}\>.
\end{equation}
Together, this yields the differential equation for the length $l$:
\begin{equation}
    \eta \frac{\mathrm{d}l}{\mathrm{d}t} = - 3 \kappa l_0^3 \left( \frac{l}{l_0}-1 \right) - 2 \gamma - 2 \gamma_p \frac{l-2 l_0}{\lambda(l)}\>.
\end{equation}

%
%
%
%

Finally, setting $\gamma=\gamma_p=\gamma_0$ yields the equation of state 
\begin{equation}
    f(l)=-3\kappa l_0^3\left (\frac{l}{l_0}-1\right )-2\gamma_0-\frac{4\gamma_0(l-2l_0)}{\sqrt{l^2-4l_0l+7l_0^2}}\>,
\end{equation}
which describes the elastic restoring force acting on the junction that is subject to a local force dipole.

The equation of state of our local model agrees well with the one describing junctional elasticity in the full tissue. This is somewhat surprising, since any junction deformation in the local model neccassarily changes the areas of all 4 cells involved~(due to limited number of degrees of freedom), whereas multiple-cells deformations in the full tissue model allow preserving cell areas. Nevertheless, provided that the local model allows significant area deformations ($\kappa=1$ in Fig.~\ref{FS2}B), the equations of state are qualitatively the same in both systems.

The complete dynamics of the local model is given by the following set of differential equations:
\begin{equation}
    \eta \frac{\mathrm{d}l}{\mathrm{d}t} = - 3 \kappa l_0^3 \left( \frac{l}{l_0}-1 \right) - 2 \gamma - 2 \gamma_p \frac{l-2 l_0}{\lambda(l)}\>,
\end{equation}
\begin{equation}
    \frac{\mathrm{d}\gamma}{\mathrm{d}t} = -\frac{1}{\tau_m} \left( \gamma - \gamma_0 \right) - \frac{\gamma}{l} \frac{\mathrm{d}l}{\mathrm{d}t}\>,
\end{equation}
\begin{equation}
    \frac{\mathrm{d}\gamma_p}{\mathrm{d}t} = -\frac{1}{\tau_m}\left( \gamma_p - \gamma_0 \right) -\frac{\gamma_p}{\lambda(l)}\frac{{\rm d}\lambda}{{\rm d}l}\frac{\mathrm{d}l}{\mathrm{d}t}\>.
\end{equation}
\section{Linear stability analysis}
\subsection{Hopf bifurcation}
We assume a small perturbation from the fixed point by setting $l=l_0+\delta l$, $\gamma = \gamma_0 + \delta\gamma$, and $\gamma_p = \gamma_0 + \delta\gamma_p$, where $\delta l,\>\delta\gamma$ and $\delta\gamma_p\ll 1$. The linearized system can be written in a matrix form as $\textup d\boldsymbol\rho/\textup dt=\underline J\boldsymbol\rho$, where $\boldsymbol{\rho}=\left (\delta l,\delta\gamma,\delta\gamma_p\right )$ and the Jacobian matrix
\begin{equation}
    \underline J = \left ( \begin{array}{ccc}
    - \frac{3\left (2\kappa l_0^3+\gamma_0\right )}{2\eta l_0} & -\frac{2}{\eta} &  \frac{2}{\eta} \\[0.2cm]
    \frac{3\gamma_0\left (2\kappa l_0^3+\gamma_0\right )}{2\eta l_0^2} & \frac{2\gamma_0}{\eta l_0} - \frac{1}{\tau_m} &  -\frac{2\gamma_0}{\eta l_0}  \\[0.2cm]
    -\frac{3\gamma_0\left (2\kappa l_0^3+\gamma_0\right )}{8\eta l_0^2} & -\frac{\gamma_0}{2\eta l_0} &  \frac{\gamma_0}{2\eta l_0} - \frac{1}{\tau_m} \\
    \end{array}\right)\>.
\end{equation}
The Jacobian has three eigenvalues: The first one, $\Lambda_1 = -1/\tau_m$, is always negative, whereas the remaining two read
\begin{equation}
    \label{S19}
    \Lambda_{\pm}=\frac{b}{2a}\left (-1\pm \sqrt{1-\frac{4ac}{b^2}}\right )\>,
\end{equation}
where $a=2\eta l_0\tau_m>0$, $b=2\left (\eta l_0+3\kappa l_0^3\tau_m-\gamma_0\tau_m\right )$, and $c=3\left (2\kappa l_0^3+\gamma_0\right )>0$. The eigenvalues are plotted in~Fig.~\ref{FS2}C.

The system is at the critical point when $\Lambda_{+}=0$; i.e., $\Lambda_{+}<0$ yields stable fixed point, whereas $\Lambda_{+}>0$ yields unstable fixed point. Since the expression in parenthesis in Eq.~(\ref{S19}) is always strictly negative, the Hopf bifurcation needs to satisfy the condition $b=0$. This gives the critical area-compressibility modulus, which reads
\begin{equation}
    \kappa^*=\frac{\gamma_0}{3 l_0^3}-\frac{\eta}{3 l_0^2 \tau_m}\>.
\end{equation}
%
\subsection{Transient oscillations}
By considering the stable system, i.e., $\kappa>\kappa^*$, we derive the condition for the existence of transient oscillations, which imply that the eiganvalues have a non-zero imaginary part. This is true when $1-4ac/b^2<0$ or
\begin{equation}
     \kappa < \frac{\gamma_0}{3l_0^3}+\frac{\eta}{3l_0^2\tau_m}+\frac{1}{3l_0^2}\sqrt{\frac{10\eta\gamma_0}{l_0\tau_m}}\>.
\end{equation}
The frequency of transient oscillations is given by $\omega=(1/2)\sqrt{4ac/b^2-1}$ and reads
\begin{equation}
    \omega=\sqrt{\frac{6\eta l_0\tau_m(2\kappa l_0^3+\gamma_0)}{(\eta l_0+3\kappa l_0^3\tau_m-\gamma_0\tau_m)^2}-1}\>.
\end{equation}
\section{First Lyapunov coefficient}
To examine the stability of the limit cycle, we calculate the first Lyapunov exponent $l_1$ at the critical point, i.e., for $\kappa=\gamma_0/(3 l_0^3)-\eta/(3 l_0^2 \tau_m)$. First, we compute eigenvectors and eigenvalues of the Jacobian $\underline J$ at the critical point and we denote by $\boldsymbol q$ the eigenvector associated to the eigenvalue $i\omega_0$ where $\omega_0 = \sqrt{-(b^2-4ac)}$. We do the same for the Jacobian transpose $\underline J^T$ and denote by $\boldsymbol p$ its eigenvector associated with eigenvalue the $-i\omega_0$, such that $\underline J^T \boldsymbol p=-i\omega_0 \boldsymbol p$. We then Taylor-expand our system around its fixed point to third order:
\begin{equation}
    \frac{d\boldsymbol\rho}{dt}=\underline F(\boldsymbol{\rho})=\underline J\boldsymbol\rho + \frac{1}{2}\boldsymbol b (\boldsymbol\rho,\boldsymbol\rho) + \frac{1}{6}\boldsymbol c (\boldsymbol\rho,\boldsymbol\rho,\boldsymbol\rho)\>,
\end{equation}
where the components of $\boldsymbol b$ and $\boldsymbol c$ read
\begin{equation}
    B_j(u,v)=\sum^3_{k,l=1}\left.\frac{\partial^2 F_j(\boldsymbol{\rho})}{\partial \rho_k \partial \rho_l}\right|_{FP}u_k v_l
\end{equation}
and
\begin{equation}
    C_j(u,v,w)=\sum^3_{k,l,m=1}\left.\frac{\partial^3 F_j(\boldsymbol{\rho})}{\partial \rho_k \partial \rho_l \partial \rho_m}\right|_{FP}u_k v_l w_m\>.
\end{equation}

The first Lyapunov coefficient $l_1$ is then given by 
\begin{align}
    l_1 &= \frac{1}{2\omega_0}  \Re [\langle \boldsymbol p,\boldsymbol c(\boldsymbol q,\boldsymbol q,\bar{\boldsymbol q})\rangle-2\langle \boldsymbol p,\boldsymbol b(\boldsymbol q,\underline J^{-1}\boldsymbol b(\boldsymbol q,\bar{\boldsymbol q}))\rangle\nonumber \\ 
    &+\langle \boldsymbol p,\boldsymbol b(\bar{\boldsymbol q},(2i\omega_0\mathbb{1}_3-\underline J)^{-1}\boldsymbol b(\boldsymbol q,\boldsymbol q))\rangle ]
\end{align}
and it turns out strictly positive for $1/\tau_m \in (0,1/l_0]$ and therefore the system never displays a stable limit cycle~(Fig.~\ref{FS2}D).
\section{Linear stability analysis of full tissue}
\begin{figure}[htb!]
    \begin{center}
    \includegraphics[width=42mm]{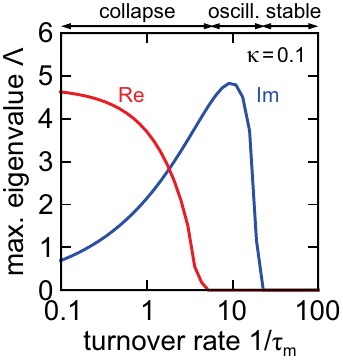}
    \caption{Maximal real and imaginary parts of the eigenvalues of the Jacobian versus $1/\tau_m$ at fixed $\kappa=0.1$.}
    \label{FS3}
    \end{center}
\end{figure}

To study the stability of the full tissue, we need to write down the system of equations in its full form. In particular, given $N_v$ vertices and $N_e$ edges, we have $2N_v$ differential equations for positions of the vertices and $N_e$ differential equations for tensions. In particular, the force exerted on vertex $i$ reads
\begin{equation}
    \boldsymbol F_i=\eta\frac{\textup d\boldsymbol r_i}{\textup dt}=-\sum_j\gamma_0\nabla_i l_{ij}-2K\sum_k\left (A_k-A_0\right )\nabla_iA_k\>,
\end{equation}
whereas the rate of change of tension at junction connecting vertices $i$ and $j$ reads
\begin{equation}
    \frac{\textup d\gamma_{ij}}{\textup dt}=-\frac{1}{\tau_m}\left (\gamma_{ij}-\gamma_0\right)-\frac{\gamma_k(\boldsymbol F_j-\boldsymbol F_i)\cdot(\boldsymbol r_j-\boldsymbol r_i)}{l_{ij}^2\eta}\>.
\end{equation}
We perturb the system around the fixed point, where the vertices assume their positions in a regular hexagonal lattice, whereas tensions on all junctions equal 1. We linearize the system, which then rereads
\begin{equation}
    \delta\boldsymbol F_i=\eta\frac{\textup d\delta\boldsymbol r_i}{\textup dt}=\sum_j\delta\boldsymbol \Lambda^l_{ij}+\sum_k\delta\boldsymbol\Lambda^A_{ik}\>,
\end{equation}
and
\begin{equation}
    \frac{\textup d\delta\gamma_{ij}}{\textup dt}=-\frac{1}{\tau_m}\delta\gamma_{ij}-\delta\Gamma_{ij}\>.
\end{equation}
Here
\begin{widetext}
\begin{equation}
    \delta\boldsymbol \Lambda^l_{ij} = \frac{\gamma_0}{l_{ij}^3}\left[ \begin{array}{c}
    \left (y_i-y_j\right )\left [\left (y_i-y_j\right )\left (-\delta x_i+\delta x_j\right )+(x_i-x_j)(\delta y_i-\delta y_j)\right ] \\ \\
    \left (x_i-x_j\right )\left [\left (y_i-y_j\right )\left (\delta x_i-\delta x_j\right )+(x_i-x_j)(-\delta y_i+\delta y_j)\right ]\\
    \end{array}\right]+
    \left[ \begin{array}{c}
    -(x_i-x_j)\delta\gamma_{ij}/l_{ij}\\ \\
    -(y_i-y_j)\delta\gamma_{ij}/l_{ij}\\
    \end{array}\right]\>,
\end{equation}
\begin{equation}
    \delta\boldsymbol \Lambda^A_{ik} = \frac{K}{2}\left[ \begin{array}{c}
    \left (y_{\nu^k(i)+1}-y_{\nu^k(i)-1}\right )\left [-\sum_{\nu\in k} (y_{\nu+1}-y_{\nu-1})\delta x_\nu+\sum_{\nu\in k}(x_{\nu+1}-x_{\nu-1})\delta y_\nu\right ]\\ \\
    \left (x_{\nu^k(i)+1}-x_{\nu^k(i)-1}\right )\left [\sum_{\nu\in k}\left(y_{\nu+1}-y_{\nu-1}\right )\delta x_\nu-\sum_{\nu\in k}\left (x_{\nu+1}-x_{\nu-1}\right )\delta y_\nu\right ]\\
    \end{array}\right]\>,
\end{equation}
and
\begin{equation}
    \delta\Gamma_{ij} = \frac{\gamma_0(\delta\boldsymbol F_j-\delta\boldsymbol F_i)\cdot(\boldsymbol r_j-\boldsymbol r_i)}{l_{ij}^2\eta}=
    \frac{\gamma_0}{l_{ij}^2\eta}\left [\delta F_j^x(x_j-x_i)+\delta F_j^y(y_j-y_i)-\delta F_i^x(x_j-x_i)-\delta F_i^y(y_j-y_i)\right ]\>.
\end{equation}
\end{widetext}
Here $\boldsymbol r_i=(x_i,y_i)$ and $l_{ij}=\sqrt{(x_i-x_j)^2+(y_i-y_j)^2}$ are positions of vertices and edge lengths, respectively, at the fixed point. Figure \ref{FS3} shows the dependence of the maximal eigenvalue of the Jacobian as a function of $1/\tau_m$ at fixed $\kappa=0.1$.
\section{Noise}
We rewrite the system of differential equations, adding the noise to the tension dynamics:
\begin{equation}
    \eta \frac{\mathrm{d}l}{\mathrm{d}t}=- 3 K l_0^3 \left( \frac{l}{l_0}-1 \right) - 2 \gamma - 2 \gamma_p \frac{l-2 l_0}{\lambda(l)}\>,
\end{equation}
\begin{equation}
    \frac{\mathrm{d}\gamma}{\mathrm{d}t} = -\frac{1}{\tau_m} \left( \gamma - \gamma_0 \right) - \frac{\gamma}{l} \frac{\mathrm{d}l}{\mathrm{d}t} + \sqrt{\frac{2 \sigma^2}{\tau_m}} \xi(t)\>,
\end{equation}
\begin{equation}
    \frac{\mathrm{d}\gamma_p}{\mathrm{d}t}= -\frac{1}{\tau_m}\left( \gamma_p - \gamma_0 \right) - \frac{\gamma_p}{\lambda(l)}\frac{{\rm d}\lambda}{{\rm d}l} \frac{\mathrm{d}l}{\mathrm{d}t} + \sqrt{\frac{2 \sigma^2}{\tau_m}} \xi_p(t)\>,
\end{equation}
where $\xi(t)$ is a Gaussian random variable with zero average and variance equal to 1. Its correlator reads
\begin{equation}
    \langle\eta(t)\eta(t')\rangle = \langle\eta_p(t)\eta_p(t')\rangle =\delta(t-t')
\end{equation}
and
\begin{equation}
    \langle\eta(t)\eta_p(t')\rangle = 0\>.
\end{equation}

The noise increases the probability for the system to escape the basin of attraction and thus expands the collapse regime in the $(1/\tau_m,\kappa)$ parameter space. Close to the boundary between the collapse and stable regimes, the system exhibits quasi-oscillations. These are instances where the noise amplifies the transient oscillations and sustains them for an indefinite time period. 

To compute the power spectrum of the quasi-oscillations, we again linearize the system around the fixed point: $l=l_0 + \delta l$, $\gamma = \gamma_0 + \delta\gamma$, and $\gamma_p = \gamma_0 + \delta\gamma_p$. 
%
%
%
The linearized system of stochastic differential equations can then be written in a matrix form as
\begin{equation}
    \frac{\mathrm{d}}{\mathrm{d}t} \boldsymbol{v} = \underline J\boldsymbol{v} + \boldsymbol{\eta}\>.
\end{equation}
Next, we write $\boldsymbol v$ in the Fourier space as $\tilde{\boldsymbol v}(\omega)=\int \boldsymbol v(t)\exp(i \omega t)$ to obtain $\left(i \omega\mathbb{1} - \underline J \right) \tilde{\boldsymbol v}(\omega) = \tilde{\boldsymbol \eta}(\omega)$. Denoting $\underline\Phi(\omega) =  i \omega\mathbb{1} - \underline J$ and solving for $\tilde{\boldsymbol v}(\omega)$ gives
\begin{equation}
    \tilde{\boldsymbol v}(\omega) = \underline\Phi^{-1}(\omega) \tilde{\boldsymbol\eta}(\omega)\>.
\end{equation}
Now we can calculate the power spectral density matrix~(PSDM) with the elements given by
\begin{equation}
P_{ij}(\omega) = \langle\tilde{v}_i(\omega)\tilde{v}_j^{\dagger}(\omega)\rangle = \sum^3_{l=1} \sum^3_{m=1} \Phi_{il}^{-1}(\omega) D_{lm} \left[{\Phi_{mj}^{-1}}\right]^{\dagger}(\omega)\>,
\end{equation}
where
\begin{equation}
D = \frac{2 \sigma^2}{\tau_m}\left( \begin{array}{ccc}
0 & 0 & 0 \\
0 & 1 & 0 \\
0 & 0 & 1
\end{array}\right)\>.
\end{equation}
The diagonal entries of the PSDM are real and coincide with the power spectra for the fluctuations, associated with the three species $l$, $\gamma$, and $\gamma_p$. In particular, the power spectrum for the junctional length fluctuations reads
\begin{equation}
    P_l(\omega) = \frac{16\sigma^2}{\tau_mg(\omega)}\left(\frac{1}{\tau_m^2}+\omega^2\right)\>,
\end{equation}
where
\begin{widetext}
    \begin{equation}
        g(\omega) = \left| -i\omega^3 + \left(\frac{\gamma_0}{l_0} - 3\kappa l_0^2 - \frac{2}{\tau_m} \right)\omega^2+i\left( \frac{1}{\tau_m^2}+\frac{\gamma_0}{2l_0\tau}+\frac{6\kappa l_0^2}{\tau_m}\right)\omega + \frac{3\gamma_0}{2 l_0 \tau_m^2}+\frac{3\kappa l_0^2}{\tau_m^2} \right|^2\>.
    \end{equation}
\end{widetext}
Note that the spectrum of the Ornstein-Uhlenbeck process is given by $P(\omega) = 2 \left(\sigma^2/\tau_m\right )/\left (\omega^2+1/\tau_m^2\right )$.
\begin{figure}[htb!]
    \begin{center}
    \includegraphics[width=42mm]{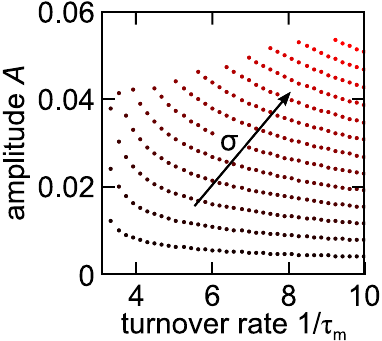}
    \caption{Amplitude of length fluctuations vs. myosin turnover rate $1/\tau_m$ for $\sigma=0.01-0.2$ (black-to-red color scheme).}
    \label{FS4}
    \end{center}
\end{figure}
\section{Correlation of quasi-oscillations}

In tissues, cell-cell junctions are physically coupled and are therefore forced to synchronize their movements. For example, junctions can lengthen in expense of the shortening of their immediate neighbors. 
\begin{figure}[b!]
    \begin{center}
    \includegraphics[]{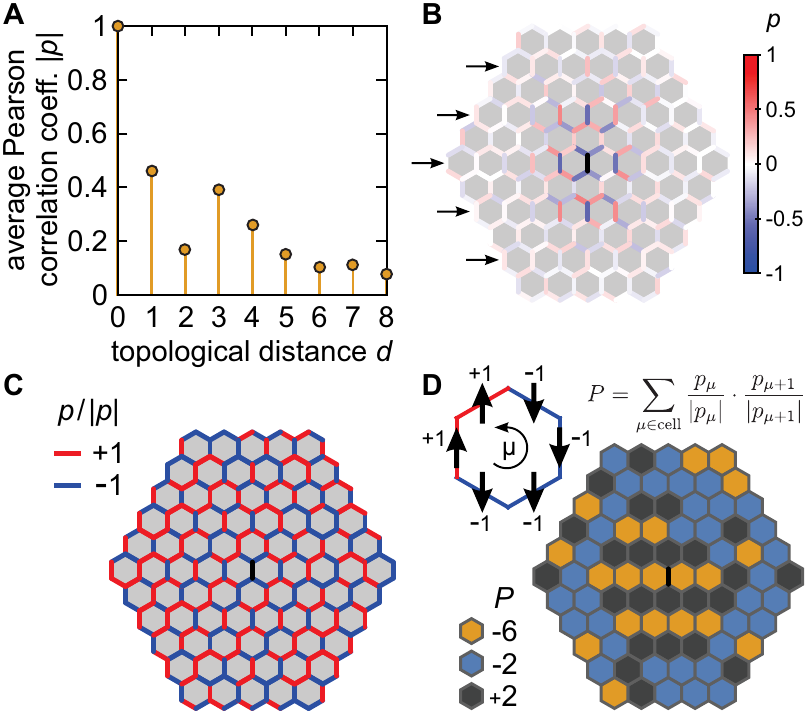}
    \caption{(A)~Pearson correlation coefficient $\left |p\right |$ vs. topological distance $d$ averaged over all junctions with equal $d$. (B)~Color map of correlation between junctions and the reference (black) junction. Black arrows point to cell chains with increased correlation strength of their parallel junctions with the reference junction. (C)~Color map of the sign of correlation between junctions and the reference (black) junction. (D)~Color map of cells' cumulative correlation of their adjacent junctions $P$. Results in all panels are calculated at $(1/\tau_m,\kappa)=(3.5,15)$.}
    \label{F3}
    \end{center}
\end{figure}
In polygonal networks, three or more junctions meet at a vertex, which prevents this pairwise relation from being simultaneously satisfied for all pairs of adjacent junctions. To explore what patterns of junctional movements this geometric frustration establishes, we simulate collective quasi-oscillations within a regular honeycomb cell tiling. We measure the Pearson correlation coefficient, $p$, between junctional lengths $l_{ij}(t)$ and the length of the reference junction $l_{\rm ref}(t)$. Not surprisingly, we find that overall, the correlation strength $\left |p\right |$ decreases with the topological distance from the reference junction $d$~(Fig.~\ref{F3}A); the topological distance $d$ is defined as the integer shortest path between a junction and the reference junction; $d= 1$ for the nearest neighbors, $d= 2$  for the next-nearest neighbors, etc. However, $\left |p\right |(d)$ is non-monotonic, which suggests a nontrivial structure of the correlation map within the cell-junctions network. Indeed, the correlation pattern exhibits two clear features: (i) correlation is stronger between junctions that are mutually parallel~(Fig.~\ref{F3}B) and (ii) chains of cells running perpendicularly to the reference junction alternate in the strength of correlation~(black arrows in Fig.~\ref{F3}B).

It is also interesting to observe how junctions synchronize their quasi-oscillations within individual cells. In particular, we focus on the sign of the correlation coefficient $p/\left | p\right |$, which tells whether the pair of junctions is correlated~($p/\left | p\right |=+1$) or anti-correlated~($p/\left | p\right |=-1$)~(Fig.~\ref{F3}C). We define cumulative correlation of adjacent junctions in a given cell by $P=\sum_{\mu\in{\rm cell}} (p_\mu/\left |p_\mu \right |)\cdot(p_{\mu+1}/\left |p_{\mu+1} \right |)$, where the index $\mu=1,...,6$. As a result of geometric frustration at the vertices, not all cells can have $P=-6$ (the case in which all pairs of adjacent junctions within a cell are anti-correlated). Interestingly, this gives rise to the formation of a local stripe-like pattern of cells, where $P$ alternates between $-6$ and $+2$, highlighting the role of mechanics in synchronization of cell deformations~(Fig.~\ref{F3}D).
\begin{figure}[htb!]
    \begin{center}
    \includegraphics[width=8.5cm]{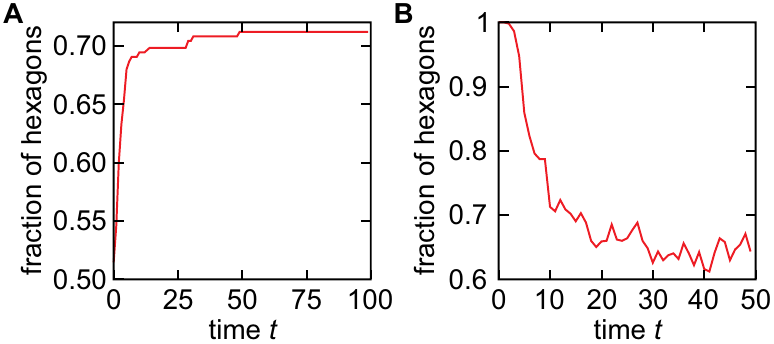}
    \caption{Fraction of hexagons versus time for a disordered initial configuration at $1/\tau_m=5.56$, $\kappa=10$, and $\sigma=0.01$~(panel A) and an ordered initial configuration at $1/\tau_m=1$, $\kappa=10$, and $\sigma=0.01$~(panel B). Tissue simulations for both cases are shown in Videos S1 and S2.}
    \label{R4}
    \end{center}
\end{figure}
%
\section{Disordered tissue samples}
To prepare disordered tissue samples, with distributed junction rest lengths, we use a pure Ornstein-Uhlenbeck scheme to describe tension dynamics:
\begin{equation}
    \dot{\Delta\gamma_{ij}}=-\frac{1}{\tau_m}\Delta\gamma_{ij}+\sqrt{\frac{2\sigma^2}{\tau_m}}\>\xi_{ij}\>.
\end{equation}
In particular, we start with a regular honeycomb lattice, which we fluidize by running simulations at $\sigma=0.5$ and $\tau_m=1$ for $t=1000$. Next, we quench the system instantaneously by setting $\sigma$ to 0 and simulate the model for $t=5000$. Due to absence of fluctuations, tensions on all junctions eventually reach 1 and spontaneous T1 transitions drive partial ordering, which stops when there are no more vanishingly short junctions.

\bibliography{apssamp}